\documentclass[sigconf,language=english]{acmart}

\renewcommand\footnotetextcopyrightpermission[1]{} 
\usepackage{tikz}
\usepackage{amsmath}
\usepackage{graphicx}
\usepackage{threeparttable}

\usepackage{multirow} 
\usepackage{listings}
\usepackage{makecell}
\usepackage{CJKutf8} 
\usepackage{tcolorbox} 
\usepackage{url}
\usepackage{booktabs,subcaption,amsfonts,dcolumn}
\usepackage{algorithm} 
\usepackage[noend]{algpseudocode} 
\usepackage{hyperref}
\usepackage{fancyhdr}
\algrenewcommand\algorithmicrequire{\textbf{Input:}}
\algrenewcommand\algorithmicensure{\textbf{Output:}}

\usepackage{pifont} 

\begin{document}

\title{Understanding Privacy Over-collection in WeChat Sub-app Ecosystem}

\author{Xiaohan Zhang}
\affiliation{
\institution{Fudan University}
\city{Shanghai}
\country{China}
}
\email{xh_zhang@fudan.edu.cn}

\author{Yang Wang}
\affiliation{\institution{Fudan University}
\city{Shanghai}
\country{China}
}
\email{20210240238@fudan.edu.cn}

\author{Xin Zhang}
\affiliation{\institution{Fudan University}
\city{Shanghai}
\country{China}
}
\email{zhangx22@m.fudan.edu.cn}

\author{Ziqi Huang}
\affiliation{\institution{Fudan University}
\city{Shanghai}
\country{China}
}
\email{huangzq21@m.fudan.edu.cn}

\author{Lei Zhang}
\affiliation{\institution{Fudan University}
\city{Shanghai}
\country{China}
}
\email{zxl@fudan.edu.cn}

\author{Min Yang}
\affiliation{\institution{Fudan University}
\city{Shanghai}
\country{China}
}
\email{m_yang@fudan.edu.cn}

\begin{abstract}

Nowadays the app-in-app paradigm is becoming increasingly popular, and sub-apps have become an important form of mobile applications. 
WeChat, the leading app-in-app platform, provides millions of sub-apps that can be used for online shopping, financing, social networking, etc. 
However, privacy issues in this new ecosystem have not been well understood. 
This paper performs the first systematic study of \underline{p}rivacy \underline{o}ver-collection in \underline{s}ub-apps (denoted as SPO), where sub-apps actually collect more privacy data than they claim in their privacy policies. 
We propose a taxonomy of privacy for this ecosystem and a framework named \emph{SPOChecker} to automatically detect SPO in real-world sub-apps.

Based on SPOChecker, we collect 5,521 popular and representative WeChat sub-apps and conduct a measurement study to understand SPO from three aspects: its landscape, accountability, and defense methods. 
The result is worrisome, that more than half of all studied sub-apps do not provide users with privacy policies. Among 2,511 sub-apps that provide privacy policies, 489 (19.47\%) of them contain SPO.
We look into the detailed characteristics of SPO, figure out possible reasons and the responsibilities of stakeholders in the ecosystem, and rethink current defense methods. 
The measurement leads to several insightful findings that can help the community to better understand SPO and protect privacy in sub-apps. 

\end{abstract}

\maketitle

\section{Introduction}
The app-in-app paradigm provides a new form of mobile applications (apps), where a \emph{super-app} acts as a host for many \emph{sub-apps} to run on. 
A super-app is often a large, popular mobile app, such as WeChat~\cite{wechat}, Line~\cite{line}, and Microsoft Teams~\cite{msteams}.
The super-apps usually provide easy-to-use development frameworks, a general running environment, and seamless distribution and integration mechanisms for sub-apps. 
More importantly, sub-apps can use the abundant resources from the super-apps, which can help them quickly acquire a large number of users. 
Therefore, more and more Internet companies, especially startups, are starting to provide services using sub-apps. 
According to a recent study~\cite{zhang2022identity}, there are even more sub-apps on WeChat than Android apps on Google Play.

\begin{figure}[!t]
    \centering
    \includegraphics[width=0.48\textwidth]{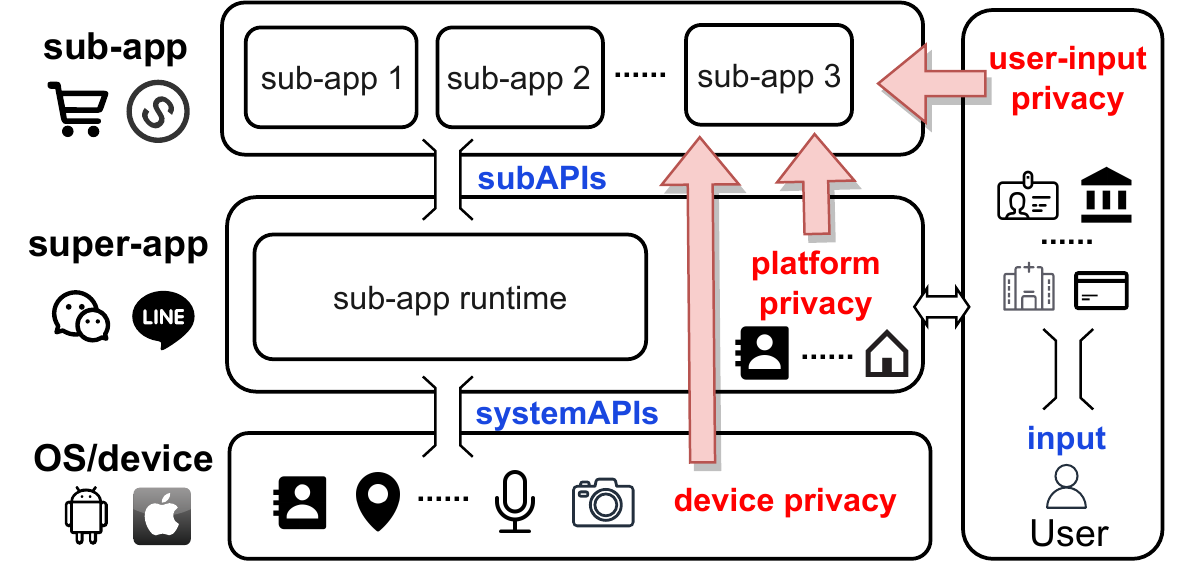}
    \caption{An Overview of the App-in-app Paradigm and the Privacy Data Sub-apps Can Access.}
    \label{fig:intro}
\end{figure}

Figure~\ref{fig:intro} illustrates the overall architecture of the app-in-app paradigm and what privacy data sub-apps can access.
The super-apps run on a mobile device and can use system APIs (\emph{systemAPIs} for short) to access device and system resources, while they provide a runtime and customized APIs (\emph{subAPIs} for short) for upper-level sub-apps. 
Therefore, depending on the source of the privacy, a sub-app can access three categories of privacy:
1) \textit{device privacy} that comes from the underlying OS and device, such as the device location, camera photos, etc;
2) \textit{platform privacy} that is produced by and stored in the super-app, such as the friend list and address information, etc; 
3) \textit{user-input privacy} that is input by the users, such as their identity information, health data, etc.
The first two categories of privacy are accessed by sub-apps by invoking subAPIs, while the last by directly interacting with the users.

One intuitive research question here is whether sub-apps \textit{over-collect} privacy data, or more specifically, \textit{do sub-apps collect more privacy data than they claim in their privacy policies}?
Unclaimed privacy collection can pose severe privacy threats to users.
For example, we find that a popular student assistant sub-app~\footnote{APPID: wx4acc7cce2e25b59d, June 2022} collects users' \textit{Bluetooth} and \textit{Device Information} by invoking subAPIs like wx.getBluetoothDevices() and wx.getSystemInfo().
However, it does not claim these collection behaviors in its privacy policy, which may cause sensitive data leakage without user awareness.

The privacy over-collection problem has been well studied in the field of mobile apps~\cite{slavin2016toward, wang2018guileak, yu2018ppchecker} and Web apps~\cite{libert2018automated, javed2020study, asif2021automated}.
However, it faces more challenges in sub-apps, resulting in previous approaches inapplicable to this new field. 
First, the development model, distribution mechanisms, and privacy policy management of sub-apps are different from those of mobile and Web apps. 
Second, privacy in sub-apps can be more complex than in mobile apps, as sub-apps can use more flexible and customized components to collect user-input data.
Third, sub-apps may access highly sensitive data that mobile and Web apps cannot. 
For example, an app on the same device as WeChat cannot access users' WeChat contact information as they are separated by system sandboxes.
However, if the app developer releases a sub-app on WeChat, it may have the ability to access contact data from WeChat.
Given the above challenges, the privacy over-collection problem in sub-apps is not well understood yet.

This paper conducts the first systematic study of sub-app privacy over-collection (denoted as SPO) in the real-world WeChat sub-app ecosystem. 
Through an in-depth study of real-world sub-apps and their development model, we give a taxonomy of the privacy available to WeChat sub-apps, which includes 37 privacy items in 3 categories. 
We then design \emph{SPOChecker} to automatically collect sub-apps and detect the over-collection of these privacy items.
Specifically, SPOChecker first collects privacy policies and all code of a sub-app, including those in on-demand loading packages, with the help of sub-app lifecycle analysis and dynamic testing.
It then uses static data flow analysis and natural language processing (NLP) to extract the sub-apps \textit{privacy collection set}, while using NLP to get the \textit{privacy claim set}. 
After that, the over-collected privacy can be obtained by calculating the set difference.

We use SPOChecker to collect 5,521 popular and representative WeChat sub-apps and conduct a measurement study from three aspects: SPO \textit{landscape}, \textit{accountability}, and \textit{defense methods}.
We find that SPO is very prevalent in WeChat sub-apps, with 15.65\% of all collection behaviors failing to inform users. 
SPO rates also vary for different privacy items and categories, with those that are harder to regulate tending to have higher SPO rates.
Templates and SDKs are heavily used and are also responsible for privacy over-collection. Furthermore, we found that the current defense mechanisms employed by WeChat can significantly improve user privacy protection, but these methods cannot cover all privacy items. 
For instance, only 34.4\% of all subAPIs that can be used to collect privacy are protected by WeChat permissions. 
Based on these findings, we can derive meaningful lessons that can help the community better protect privacy in this emerging field.

In summary, this paper makes the following contributions:
\begin{itemize}
    \item We provide a taxonomy of privacy in WeChat sub-apps and conduct the first systematic study on privacy over-collection in this ecosystem.
    
    \item We propose SPOChecker, a tool combining static analysis, dynamic testing, and NLP techniques to automatically collect sub-apps and detect SPO in real-world sub-apps.
    
    \item We make a large-scale real-world measurement and demystify the landscape, accountability, and defense of SPO, which leads to several useful findings that can help the community better protect user privacy in this emerging field. 
    
\end{itemize}

\textbf{Organization}. 
The rest of this paper is organized as follows. 
In \S\ref{section:background}, we present the background and statement of the SPO problem. 
Detailed steps and evaluation of SPOChecker are proposed in \S\ref{section:SPOChecker}. 
In \S\ref{section:measurement}, we build a large-scale dataset and conduct a measurement on SPO by answering several research questions. 
Discussions are proposed in \S\ref{section:discussion}, and related works in \S\ref{section:related work}.
Finally we conclude the paper in \S\ref{section:conclusion}.

\section{The SPO Problem}~\label{section:background}
This section provides an overview of WeChat sub-apps, presents our taxonomy of privacy in sub-apps, and discusses the problem statement to clarify the SPO problem.

\subsection{The Structure of WeChat Sub-apps}

\begin{figure}[htbp]
    \centering
    \includegraphics[width=0.35\textwidth]{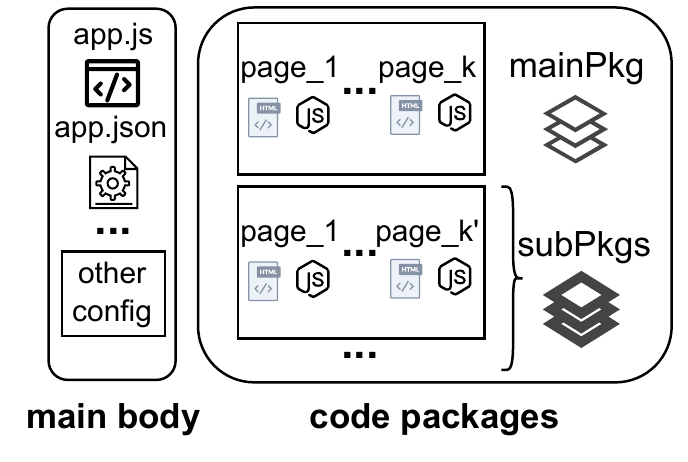}
    \caption{The Structure of WeChat Sub-apps.}
    \label{fig:subapp-structure}
\end{figure}

As previously shown in Figure~\ref{fig:intro}, a super-app provides a runtime, usually based on  WebView~\cite{googlewebview, wechatwebview} for sub-apps to run on. 
Thus sub-apps are generally a special type of Web app. 
Figure~\ref{fig:subapp-structure} shows the structure of a WeChat sub-app after abstraction without loss of correctness. 

\textbf{WeChat sub-app structure}.
A WeChat sub-app consists of two parts, the main body, and several code packages. 
The main body contains an instance file (app.js) and some global configuration files (app.json, etc.).
In app.js, the sub-app registers its instance and binds lifecycle callbacks and event listeners. 
In app.json and other configuration files, the sub-app specifies its global configurations, including its basic information, overall style, network settings, etc. 
Furthermore, the list of all pages of a sub-app is also stated in the configuration files.

\textbf{Code packages and on-demand loading}.
A sub-app contains one main code package (\emph{mainPkg} for short) and zero or more supplementary code packages (\emph{subPkgs} for short). 
Each code package has several ``pages'', where a page is the basic component of a sub-app, like an ``Activity'' in Android apps.
The entry point of a sub-app, or entry page, is stored in the mainPkg. 
To optimize loading efforts, WeChat adopts on-demand loading of subPkgs.
That is, only the mainPkg is loaded when opening a sub-app, while subPkgs are dynamically loaded when users visit the corresponding pages.

\textbf{Sub-app pages and subAPIs}.
Each page consists of a render layer and a logic layer, which are composed of HTML-like files and JavaScript files respectively. 
The render layer is responsible for displaying the page to users, while the logic layer performs data updating and networking operations. 
As previously shown in Figure~\ref{fig:intro}, sub-apps can use JavaScript to invoke subAPIs provided by super-apps, to access and obtain various resources.

\subsection{Privacy in Sub-apps: A Taxonomy}~\label{subsec:a taxonomy}

The structure of WeChat sub-apps determines that they can access three types of privacy, \emph{device privacy} and \emph{platform privacy} through subAPIs, and \emph{user-input privacy} through UI interactions. 
However, to the best of our knowledge, no previous work has systematically studied the complete list of privacy items a sub-app can access.

\begin{table*}
\centering
\caption{Taxonomy of Privacy in WeChat Sub-apps.}~\label{table:taxonomy}
\scalebox{0.79}{
    \begin{tabular}{llp{15.5cm}}
    \toprule
    \textbf{Category} & \textbf{Privacy Items} & \textbf{SubAPI \& Keywords} \\
    \midrule
    \multirow{15}{*}{\begin{tabular}[c]{@{}l@{}}
    \textbf{Device privacy}\\
    (15 items)
    \end{tabular}}& device information &  getSystemSetting(), getSystemInfo(), getDeviceInfo(), startLocalServiceDiscovery(), getBatteryInfo(), getScreenBrightness()\\
    
     & photographic image & \begin{tabular}[c]{@{}l@{}}
     shareVideoMessage(), chooseImage(), chooseVideo(), chooseMedia(), createLivePusherContext.sendMessage(), \\
     showShareImageMenu()
     \end{tabular}\\
     
     & file & \begin{tabular}[c]{@{}l@{}}
      shareFileMessage(), uploadFile(), chooseMessageFile(), getSavedFileList(), getSavedFileInfo(), getFileInfo(), \\
      getSavedFileList(), readFile(), readCompressedFile(), openDocument()
      \end{tabular}\\
     
     & location information & \begin{tabular}[c]{@{}l@{}}  createMapContext.fromScreenLocation(), createMapContext.getCenterLocation(), createMapContext.toScreenLocation(),\\
     chooseLocation(), startLocationUpdateBackground(), openLocation(), onLocationChange(), getLocation(), choosePoi(),\\
     startLocationUpdate()
     \end{tabular}\\
     
     & screenshot & createLivePusherContext.snapshot() \\
     
     & camera & \begin{tabular}[c]{@{}l@{}} startPreview(), switchCamera(), startRecord(), onCameraFrame(), takePhoto(), CameraFrameListener.start(), scanCode(),\\
     getVKFrame(), subscribeVoIPVideoMembers(), join1v1Chat(), onVoIPVideoMembersChanged(), initFaceDetect(), \\
     faceDetect(), createVKSession.start()\end{tabular}\\
     
     & recording & \begin{tabular}[c]{@{}l@{}} startRecord(), getRecorderManager(), setEnable1v1Chat(), joinVoIPChat(), subscribeVoIPVideoMembers(), join1v1Chat(),\\
     onVoIPVideoMembersChanged() \end{tabular}\\
     
     & biometric information & startSoterAuthentication(), checkIsSupportSoterAuthentication(), checkIsSoterEnrolledInDevice(), initFaceDetect() \\
     
     & bluetooth & \begin{tabular}[c]{@{}l@{}} isBluetoothDevicePaired(), openBluetoothAdapter(), onBluetoothDeviceFound(), onBluetoothAdapterStateChange(),\\
     readBLECharacteristicValue(), getConnectedBluetoothDevices(), getBluetoothDevices(), getBluetoothAdapterState(),\\
     writeBLECharacteristicValue(), startBluetoothDevicesDiscovery(), onBLEConnectionStateChange(), getBLEDeviceRSSI(),\\
     createBLEConnection(), getBLEDeviceServices(), onCharacteristicReadRequest(), getBLEDeviceCharacteristics(),\\
     writeCharacteristicValue(), startAdvertising(), startBeaconDiscovery(), onBeaconUpdate(), onBeaconServiceChange(),\\
     
     makeBluetoothPair(), getBeacons(), onBLEPeripheralConnectionStateChanged(), faceDetect()
     \end{tabular}\\
     
     & NFC & \begin{tabular}[c]{@{}l@{}}
     getNFCAdapter(), getIsoDep(), getMifareClassic(), getMifareUltralight(), getNdef(), getNfcA(), getNfcB(), getNfcF(), \\getNfcV(), startHCE(), sendHCEMessage(), onHCEMessage(), getHCEState() \end{tabular}\\
     
     & clipboard data & setClipboardData(), getClipboardData() \\
     
     & network information & \begin{tabular}[c]{@{}l@{}}onNetworkWeakChange(), onNetworkStatusChange(), getNetworkType(), getLocalIPAddress()\\
     startWifi(), setWifiList(), onWifiConnected(), onGetWifiList(), getWifiList(), getConnectedWifi(), connectWifi()\end{tabular} \\
     
     & calling information & makePhoneCall() \\
     
     & sensor data & \begin{tabular}[c]{@{}l@{}} startAccelerometer(), onAccelerometerChange(), startCompass(), onCompassChange(), startDeviceMotionListening(),\\
     onDeviceMotionChange(), startGyroscope(), onGyroscopeChange() \end{tabular}\\
     
     & contact information & chooseContact() \\
     \midrule
     
     
    \multirow{5}{*}{\begin{tabular}[c]{@{}l@{}}
    \textbf{Platform privacy}\\
    (5 items)
    \end{tabular}}
    & contact information & getPhoneNumber()  \\
     & property information & chooseInvoiceTitle(), chooseInvoice() \\
     & activity information & getWeRunData() \\
     & driving information & chooseLicensePlate() \\
     & location information & chooseAddress() \\
     \midrule
     
    \multirow{17}{*}{\begin{tabular}[c]{@{}l@{}}
    \textbf{User-input privacy} \\
    (17 items)
    \end{tabular}} & name & "name", "contact person", "consignee" \\
     & gender & "gender" \\
     & age & "age", "date of birth" \\
     & ethnic group & "ethinic group" \\
     & nationality & "nationality" \\
     & political or religious & "political or religious views" \\
     & marriage situation & "marriage situation" \\
     & driving information & "vehicle number", "license plate number", "vehicle information", "driver’s license", "driving license" \\
     
     & location information & \begin{tabular}[c]{@{}l@{}}"address", "region", "location", "gate number", "residence", "neighborhood", "venue", "community", "street", "floor", \\
     "postcode", "longitude and latitude", "longitude", "latitude", "residence permit", "resident Committee", "district and county", \\
     "country", "place of attribution", "place of departure", "place of destination", "accommodation information", \\
     "household registration", "province", "city", "village", "whereabouts", "destination"
     \end{tabular} \\
     
     & device information & "device name" \\
     & contact information & "phone number", "contact information", "contact number", "email address" \\
     
     & identity information & \begin{tabular}[c]{@{}l@{}}
     "ID card", "ID number", "the front side of ID card", "the back side of ID card", "business license", "certificate number",\\
     "medical insurance card number", "patient identification number", "health card", "health code", "certificate", "certificate ID"
     \end{tabular}\\
     
     & physiological and health & "pathological information", "illness", "weight", "height" \\
     & online identity information & "password", "account", "Alipay account", "payment password" \\
     & activity information & "flight number", "boarding pass", "flight date" \\
     
     & property information & \begin{tabular}[c]{@{}l@{}}"bank card", "bank name", "bank account", "bank of account", "cardholder", "credit card", "deposit card", "financial account",\\
     "invoice title", "tax number", "invoice", "property information"
     \end{tabular} \\
     
     & education and career & \begin{tabular}[c]{@{}l@{}}"educational experience", "academic degree", "major", "biography", "work unit name", "work unit", "school name", \\
     "occupation", "employee number", "position", "company name", "legal person", "enterprise", "social credit code", \\
     "work experience", "title", "contract", "student number"
     \end{tabular}\\
     \bottomrule
    \end{tabular}
}
\begin{tablenotes}
\footnotesize
\item{* Privacy keywords in user-input privacy are translated from Chinese and one Chinese word can be translated into multiple synonyms in English. E.g. "\begin{CJK}{UTF8}{gbsn}生日\end{CJK}" can be translated into "birthday" or "date of birth", and we only list one of them here.}
\end{tablenotes}
\end{table*}

Therefore, in this paper, we propose the first taxonomy of privacy a sub-app can access in the WeChat sub-app ecosystem. 
To set up the taxonomy, we carefully study the official API document~\cite{subapidocument}, investigate real-world sub-apps, and refer to authoritative privacy standard~\cite{gb2020information,mccallister2010guide} and prior work~\cite{kokolakis2017privacy}.
We collect sensitive items about personal information (PI) from the above official standards and filter out those that cannot be obtained in WeChat sub-apps. 
Then we categorize the remaining privacy items according to the property of PI referring to the authoritative classification mentioned above.

Finally, we summarize 37 privacy items in three categories, as shown in Table~\ref{table:taxonomy}.
Note that some privacy items may appear in multiple categories because they can be collected in different ways. 
For example, invoice data can be collected by calling subAPI wx.chooseInvoice(), or from a user-input component.
As a result, there are 29 distinct privacy items in this taxonomy.
In the rest of this paper, we use a suffix to distinguish privacy items from different categories, e.g. \emph{contact\_d}, \emph{contact\_p}, and \emph{contact\_u} represent contact privacy from device, platform, and user-input respectively.

Especially, platform privacy is a novel privacy category unique to sub-app ecosystems.
As super-apps like WeChat often have very sensitive platform data, such as social and financial information, the over-collection of this kind of data may cause more serious privacy and security risks than previous mobile platforms.

Table~\ref{table:taxonomy} also lists all the subAPIs and privacy keywords that can be used by sub-apps to obtain these privacy items.
SPOChecker relies on this table to find all privacy collection behaviors of sub-apps.

\subsection{Problem Statement}

In this paper, we study the SPO problem by calculating a set $\boldsymbol{S_{spo}}$ for each sub-app using the following equation:
\begin{displaymath}~\label{def:SPO}
    \boldsymbol{S_{spo}} = \boldsymbol{S_{collect}} - \boldsymbol{S_{claim}}
\end{displaymath}
where $\boldsymbol{S_{collect}}$ is the set of the privacy items it collects and $\boldsymbol{S_{claim}}$ it claims in privacy policies.
The privacy items are those defined in our taxonomy in Table~\ref{table:taxonomy}.
Note that a previous work~\cite{lu2020demystifying} shows that a sub-app may illegally access more resources by privilege escalation and invoking hidden platform APIs. 
We do not consider this kind of attack and only focus on privacy items that a sub-app can access using regular and legal methods. 

The key challenge here is to accurately and completely get $S_{collect}$ for a sub-app. 
If a sub-app only uses the privacy data locally without sending them to its server, it should not be seen as a collection behavior. 
Therefore, we should carefully define the source and sink points of privacy data, as well as track the data propagation in the sub-app code.

\section{Detecting SPO}~\label{section:SPOChecker}

\begin{figure*}[htbp]
    \centering
    \includegraphics[width=0.9\textwidth]{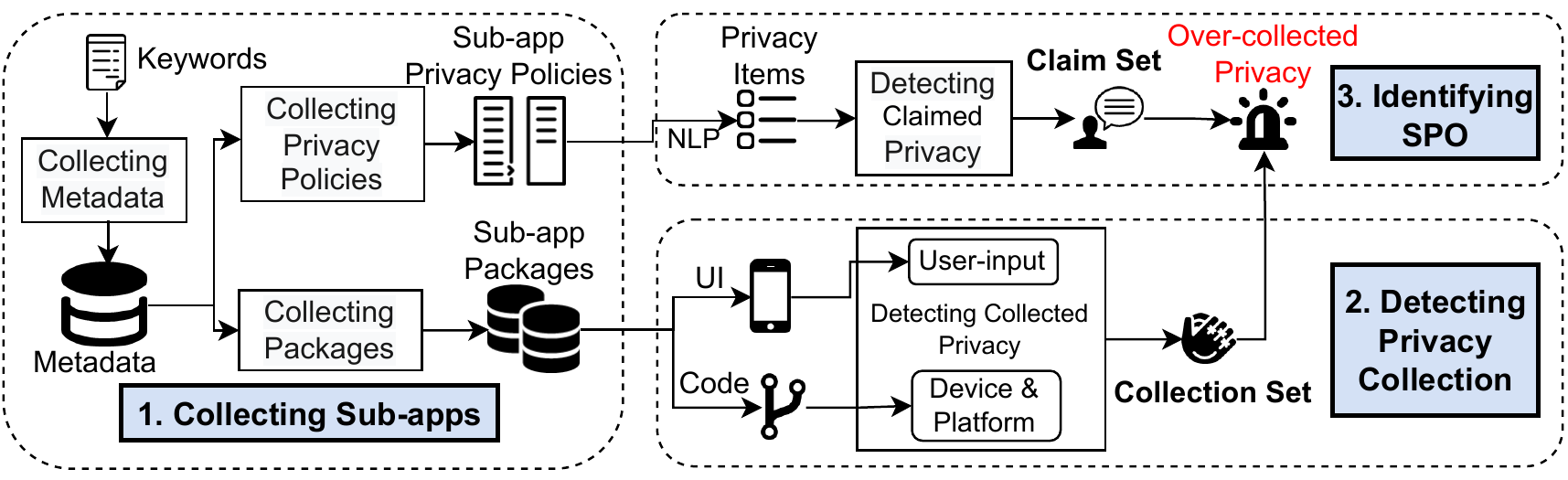}
    \caption{The Overall Workflow of SPOChecker.}
    \label{fig:SPO_detection}
\end{figure*}

In this section, we describe the workflow of SPOChecker on how it automatically collects sub-apps and identifies SPO.
As shown in Figure~\ref{fig:SPO_detection}, SPOChecker has three main steps.

\subsection{Step 1: Collecting Sub-apps}~\label{subsection: collecting}

The first step is to collect sub-apps, including all their code and privacy policies. 
However, unlike Android and iOS which have app markets, there is no such market for WeChat sub-apps.
A prior work~\cite{zhang2021measurement} utilizes the search interface in WeChat to collect sub-apps, but it can only collect \emph{mainPkgs} while neglecting the \emph{subPkgs} of these sub-apps.
Also, it does not collect any privacy policies.
In this paper, we start with the same idea, but we improve the collection by including all code packages and privacy policies.
More specifically, we use dynamic testing to automatically trigger the on-demand loading of all code packages, and then dump them by hooking certain WeChat functions. 
We describe the detailed process below.

\textbf{Collecting metadata.} 
We first determine what sub-apps should be collected by obtaining sub-app metadata, including its ``APPID'', ``developer'', ``category'', ``recently used'', etc. 
To achieve this, we intercept the searching API~\footnote{https://mp.weixin.qq.com/wxa-cgi/innersearch/mmbizwxasearchapp, Jun 2022} and then feed different keywords into this API and then extract the metadata from search results. 
Note that we can get more relevant keywords from this metadata, and then we repeat the previous step until the number of sub-apps ceases to increase. 
Specifically, to make the collection of sub-apps as diverse as possible, we collected 191 app category keywords from major app markets~\cite{googleplay}, and 560 top-ranked sub-apps and app names from major app data ranking websites~\cite{alading, chandashi}, as the initial keyword list.

\textbf{Collecting code packages.}
After we get the metadata of a sub-app, we use dynamic testing to download all its code packages by hooking the WeChat app. 
First, we use UIAutomater~\cite{uiautomator} to let the WeChat app load a sub-app based on its APPID, when the main body and mainPkg of the sub-app can be dumped. 
Then we parse the configuration files to extract the route to all subPkgs, force WeChat to trigger on-demand loading, and dump these subPkgs.
In this way, we can collect all code of a sub-app.

\textbf{Collecting privacy policies.}
Both Android and iOS app markets require developers to display privacy policies on the app downloading page. 
However, there are no such rules for sub-apps, so we need to locate the privacy policies in sub-apps. 
By examining real-world sub-apps, we find privacy policies may be provided as text files or links that needed to be redirected. 
Consequently, we choose to collect privacy policies in a dynamic way. 
Given the observation that privacy policies are always displayed on certain pages to increase user awareness, we trigger all the pages that contain privacy-related keywords. 
Then we parse the page layout, locate the component containing keywords, simulate clicking into the detail page of the privacy policy, and extract the full privacy policy text. 
Furthermore, when privacy policies are presented as images or PDFs, we utilize OCR tools~\cite{ocr} to extract the text.

\subsection{Step 2: Detecting Privacy Collection}~\label{subsection:detecting collection}

This subsection outlines how SPOChecker identifies the collection of privacy data for three sub-app categories.
Specifically, we first describe how SPOChecker locates the \textit{Source} of all privacy collection behavior in the first three parts of this subsection, and then describe how SPOChecker decides whether the data is \textit{Sink}ed to the sub-apps servers.

\subsubsection{Platform Privacy}~\label{subsubsection: pltapi}
To locate what platform privacy a sub-app locally collects is relatively straightforward, as WeChat subAPI documents~\cite{subapidocument} have listed all subAPIs that can collect platform privacy, as listed in Table~\ref{table:taxonomy} of this paper.
Therefore, SPOChecker finds all invocation of platform device collecting subAPIs and then finds the success and complete callbacks, or more specifically the returned data of these callbacks, and marks them as the \textit{Source} points.

\subsubsection{Device Privacy}~\label{subsubsection: deviceapi}
Unlike platform privacy, there is no existing work that can map all device privacy-collecting subAPIs to corresponding privacy items. 
The WeChat API documents provide descriptions for each subAPI but do not explicitly specify what device privacy a subAPI collected.
WeChat provides a large number of subAPIs, i.e. 973 subAPIs in official documents~\cite{subapidocument}, thus manually mapping them to privacy items is infeasible. 
A prior work~\cite{lu2020demystifying} utilizes fuzzing to find which systemAPIs a subAPI invocation may eventually reach, thus inferring the system resources a subAPI can access.
However, it only considers the invocation of a single subAPI during each fuzzing test, while neglecting the dependency between subAPIs, thus may fail to find some privacy-collecting subAPIs.
Furthermore, it mainly focuses on privilege escalation subAPIs.
As a result, it cannot meet the demand of our work.

\begin{table}[htbp]
\centering
\caption{New SubAPIs We Found That Can Be Used to Collect Device Privacy.}~\label{table:new privacy subAPIs}
\resizebox{\linewidth}{!}{
    \begin{tabular}{l l l}
    \toprule
    \textbf{SubAPI} & \textbf{SystemAPI } & \textbf{Privacy Items} \\
    \midrule
    wx.getFileSystemManager & StorageManager.getVolumeList & file\\
    wx.onWifiConnected & WifiManager.getConnectionInfo & network\\
    wx.getConnectedWifi & WifiManager.getConnectionInfo & network\\
    wx.connectWifi & \makecell[l]{WifiManager.getConnectionInfo\\ConnectivityManager.requestNetwork} & network\\
    wx.chooseContact & ContentResolver.registerContentObserver & contact\\
    wx.getClipboardData & ClipboardManager.getPrimaryClip & clipboard data\\
    wx.startAccelerometer & SensorManager.registerListener & sensor data\\
    wx.onAccelerometerChange & SensorManager.registerListener & sensor data\\

    \bottomrule
    \end{tabular}
}
\end{table}

In this paper, we optimize the fuzzing method proposed in~\cite{lu2020demystifying} by considering control and data flow dependencies between subAPIs to generate more effective testing cases. 
Our method has the following steps: 
1) For all 973 WeChat subAPIs, we exclude those that are unrelated to device privacy, such as subAPIs used for UI layout and styles, and subAPIs that cannot be used to get meaningful data, such as \textit{wx.closeSocket()} and \textit{wx.getRandomValue()}.
After this step, the number of subAPIs to analyze is reduced from 973 to 153.
2) We refer to sample code in API documents and real-world sub-apps to generate test cases for the above 153 subAPIs, where 55 (36\%) of them have control or data flow dependency with other subAPIs, so we add necessary dependencies to generate high-quality testing cases.
3) We fuzz these subAPIs and monitor what systemAPIs they can reach by monitoring systemAPIs using frida~\cite{frida}.
If a subAPI invocation finally gets data from systemAPIs that collect device privacy, we manually check whether it collects any device privacy.
Here the list of sysytemAPIs that can collect device privacy is generated by combining APIs in several classical studies~\cite{aafer2018precise, dawoud2021bringing, backes2016demystifying}.

In summary, we identify 125 WeChat subAPIs that can be used to collect 15 types of privacy items, as shown in Table~\ref{table:taxonomy}.
We also build a complete mapping (also open-sourced in our repo) from subAPI invocation to privacy item collection.
Note that this mapping is generated off-line, and once generated it can be used to detect device privacy collection in all sub-apps. 
Specifically, Table~\ref{table:new privacy subAPIs} lists some device privacy-collecting APIs that cannot be found by previous method~\cite{lu2020demystifying}.

As done in handling platform privacy, we mark the success and complete callbacks of these 125 subAPIs as the \textit{Source} points of device privacy collection.

\subsubsection{User-input Privacy}~\label{subsubsection:uip}
Sub-apps can also collect user-input privacy, known as UIP in previous work~\cite{nan2015uipicker}, through UI interactions with users.
Previous work~\cite{nan2015uipicker, huang2015supor} mainly studies a limited set of UIP on mobile apps.
However, the forms of UIP in sub-apps are more diverse, because sub-apps, which are written in Web languages, can customize the input components more flexibly compared to mobile apps on Android or iOS. 
As a result, the widely used customized input components, as well as dependencies between these Web components, bring new challenges to identifying UIP in sub-apps.

\begin{figure}[htbp]
    \centering
    \includegraphics[width=0.4\textwidth]{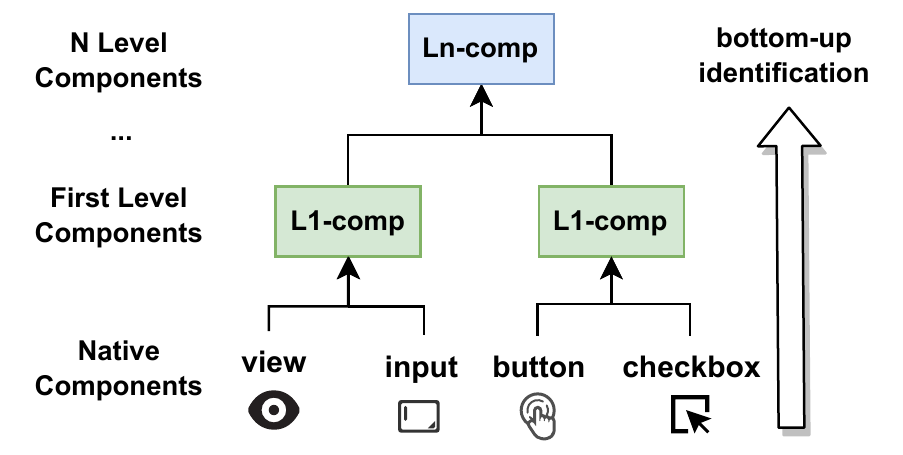}
    \caption{Customized UIP-related Components in Sub-apps and the Bottom-up Idea.}
    \label{fig:customized component}
\end{figure}

To address these challenges, we propose a bottom-up identification method to recognize customized UIP-related components.
First, we extract all customized UIP-related components that depend on native components, called first-level components. 
After that, second-level components, i.e., customized components that depend on native and first-level components, are identified. 
By parsing the hierarchy bottom-up, we eventually find all UIP-related components. 
Then we can locate all UIPs by matching sensitive privacy words in their texts.

\begin{table}[htbp]\footnotesize
\centering
\setlength{\tabcolsep}{3pt}
\caption{Data Binding Patterns for Native Components.}~\label{table:uip_component}
    \begin{tabular}{lll}
    \toprule
    \textbf{Type}
     & 
    \begin{tabular}{l}
    \textbf{Native Components}
    \end{tabular}
     & 
    \begin{tabular}{l}
    \textbf{Binding Pattern} 
    \end{tabular}
     \\
    \midrule
    \textbf{Input} 
    & 
    \begin{tabular}{l}
    editor, input, textarea 
    \end{tabular}
    & 
    \begin{tabular}{l}
    bindinput, bindconfirm
    \end{tabular}
    \\
    \addlinespace[0.3em]

    \textbf{Process} 
    & 
    \begin{tabular}{l}
    checkbox, checkbox-group, \\
    picker, picker-view, 
    radio, \\ radio-group, 
    slider, switch, form 
    \end{tabular}
    & 
    \begin{tabular}{l}
    bindchange, catchsubmit, \\
    bindcolumnchange, \\
    bindchanging
    \end{tabular}
    \\ 

    \bottomrule
    \end{tabular}
\end{table}

We then map the UI components to their underlying processing code, specifically, which variables in the logic layer receive any collected privacy data directly from the UI components.
The WeChat subAPI documents~\cite{wechat} have listed all the data binding patterns for the native components, which are collated and listed in Table~\ref{table:uip_component}.
Note that the basic two types of native components,  i.e. ``Input'' and ``Process'',  have different binding patterns, and higher-level components inherit the patterns of the native components. 
We refer to these observations to get the bindings for all UIP-related components. 
As an example, in `\textless picker bindchange=``handleCityChange'' ... \textgreater', SPOChecker will locate the \textit{handleCityChange} function in the logic layer, whose parameters will be marked as the \textit{Source} points.

\subsubsection{Data Flow Analysis}~\label{subsubsection:DFA}
After finding all the \textit{Source} points of privacy collection, SPOChecker then conducts a static data flow analysis on the logic layer to find whether the data is \textit{Sink}ed to the sub-app's server.

There are several works that are related to static analysis on JavaScript code. 
For example, SAFE~\cite{SAFE} and JSAI~\cite{jsai} use static analysis to find JavaScript syntax and type errors, while TAJS~\cite{tajs}, JSFlow~\cite{jsflow} and JSPrime~\cite{jsprime} are proposed to do data flow analysis on JavaScript code. 
However, WeChat sub-apps differ from general JavaScript programs due to the existence of the sub-app framework, lifecycle management, and event handling mechanisms, etc. 
As a result,  existing tools cannot be directly applied to WeChat sub-apps.

Therefore, in SPOChecker we develop a customized static data flow analysis for WeChat sub-apps, which tracks the data transmission from the \textit{Source} to the \textit{Sink}. 
By referring to existing works~\cite{odgen, nodest, li2021detecting} that use taint analysis to locate JavaScript vulnerabilities, we customize our tool to solve the unique problems in the context of sub-apps. 
The main steps are described below.

\textbf{Building call graph.} 
Given the code of the sub-apps, we first generate the abstract syntax tree (AST) based on Esprima~\cite{esprima}.
As shown in Figure~\ref{fig:subapp-structure}, a sub-app is made of an \textit{App} instance (app.js) and several \textit{Page}s, which all have several key global functions and variables.
According to the sub-app and page registration framework specified by WeChat, functions and variables in the app instance and each page may have pre-defined lifecycles and callbacks. 
Therefore, we set up a main page loading function as the \textit{dummyMain} method in Flowdroid~\cite{flowdroid}, and add these implicit calls to the generated AST to build a complete call graph. 

\textbf{Tracking the data flow. } 
SPOChecker then tracks the data flow along the call graph to find whether the tainted source is sinked to the sub-apps server. 
Specifically, SPOChecker tracks all intra-procedure and all direct inter-procedure data transmissions.
Complicated inter-procedure data transmission, such as first storing privacy data in a file and then uploading it from another process, is not considered in this step. 
However, our evaluation (\S~\ref{subsec:evaluating}) shows that such complex data transmissions are rare in sub-apps.

\begin{table}[htbp]
\centering
\caption{Sink SubAPIs. }~\label{table:sink_api}
\scalebox{0.86}{
\resizebox{\linewidth}{!}{
    \begin{tabular}{ll}
    \toprule
    \textbf{Category}
     & 
    \begin{tabular}{l}
    \textbf{subAPI}
    \end{tabular}
     \\
    \midrule
    \textbf{Upload} & 
    \begin{tabular}{l}
    wx.uploadFile(), wx.sendMessage()
    \end{tabular}
    \\
    \addlinespace[0.3em]
    
    \textbf{Request} & 
    \begin{tabular}{l}
    wx.request(), wx.sendSocketMessage(), \\
    SocketTask.send(), wx.createTCPSocket(), \\
    TCPSocket.write(), wx.createUDPSocket(), \\
    UDPSocket.send(), UDPSocket.write()
    \end{tabular}
    \\ 
     
    \bottomrule
    \end{tabular}
}
}
\end{table}

\textbf{Checking the sinks.} 
As shown in Table~\ref{table:sink_api}, SPOChecker focuses on subAPIs that can be used to upload files or resources (the ``Upload'' type) and send requests (the ``Request'' type). 
Note that there may be some network libraries based on these basic subAPIs, and SPOChecker can track them into these libraries. 
If these \textit{Sink} points are reached, SPOChecker then finds a privacy collection behavior that sends privacy to the sub-app's server.

After the above steps, SPOChecker can get the privacy collection set of the analyzed sub-apps.

\subsection{Step 3: Identifying SPO}~\label{subsection: identifying SPO}
We then identify from the privacy policy what privacy sub-app claims, as sub-apps in our dataset are for the Chinese market so their privacy policies are in Chinese.
SPOChecker takes the following steps to identify privacy statements in them:

\textbf{Generating keyword list.} 
We first use \textit{Named Entity Recognition} to recognize privacy keywords in privacy policies. Inspired by~\cite{andow2019policylint}, we adapt the BERT model~\cite{devlin2018bert} in Chinese to the privacy policy domain and generate a keyword list for privacy items. 
Note that we have tested different NLP pre-trained models~\cite{devlin2018bert, lan2019albert} and the BERT model performs the best.
Additionally, we refine the list by adding synonyms and manually found keywords to make sure its completeness.

\textbf{Identifying candidate sentences.}

After getting the complete keyword list, we then use NLP tools to process the privacy policies. 
Before matching keywords, we first tokenize each sentence into words.
We then add those sentences that contain at least one keyword to the candidate list.

\textbf{Confirming collection claim. }
We then conduct non-collective statement recognition and negative sentiment analysis to exclude sentences such as ``please call our contact number'' or ``we do not collect your address''.
Specifically, we inspect the verb to exclude statements without collecting behavior and propagate the negative sentiment in the sentence by switching the sentiment label of each matched keyword to filter out non-collective statements.

Having acquired the complete sub-app privacy collection set and privacy claim set, we bring SPO to light by computing the difference set between the former and the latter.

\subsection{Evaluating SPOChecker}~\label{subsec:evaluating}

We evaluate the performance of SPOChecker in locating privacy policies and detecting privacy over-collection behaviors in sub-apps.

\textbf{Locating privacy policies}. 
We selected a random sample of 1,000 sub-apps from our dataset (\S~\ref{subsection: dataset}) for manual verification. 
Using SPOChecker, we identified 445 sub-apps with valid privacy policies and 518 sub-apps lacking privacy policies. 
Next, two security experts were asked to review the 445 identified privacy policies. For the 518 sub-apps lacking privacy policies, we conducted manual searches by dynamically running the sub-apps in order to identify any present privacy policies.
Our manual verification confirmed the accuracy of SPOChecker's findings.
In addition, SPOChecker identified 37 sub-apps with invalid privacy policies, which we further manually checked and found that the privacy policies were indeed inaccessible, such as the absence of responses after a user clicks or a blank detail page.
Overall, our experiment demonstrates the effectiveness of SPOChecker in accurately and comprehensively identifying privacy policies for sub-apps.

\textbf{Detecting SPO}.
We randomly selected 100 sub-apps with valid privacy policies from the above samples and had two security experts dynamically run these sub-apps and manually inspect their code, to construct the ground truth of privacy collection behaviors for them.
Overall, these 100 sub-apps have 283 privacy collection behaviors, while 55 of them are SPO.

\begin{table}[htbp]
\centering
\caption{Manual Check Result. }
\label{table:manualcheck}
\scalebox{1}{
    \begin{tabular}{ccc}
    \toprule
      & \textbf{\# Collected Privacy} & \textbf{\# SPO}  \\
    \midrule
    \textbf{Ground Truth} & 283 & 55  \\
    \textbf{SPOChecker} & 280 & 50  \\
    \bottomrule
    \end{tabular}
}
\end{table}

We then apply SPOChecker on these 100 sub-apps, and it successfully detects 280 (98.94\%) privacy collection behaviors and 50 (90.91\%) SPO behaviors, as shown in Table~\ref{table:manualcheck}.
Since SPOChecker is designed to be conservative as discussed in \S~\ref{subsubsection:DFA}, it does not report any false positives of both collection and over-collection behaviors.

We conduct further analysis to determine the reasons for the false negatives. 
The primary reason for the 3 false negatives in identifying privacy collection behavior (thus also for SPO detection) is the inability of static analysis to recognize complex data transmission.
For instance, we observe that the sub-app stores collected user email as page data (data.email) but does not immediately upload it to its server within the current process. 
Instead, it retrieves the email information from another process before uploading it. 
However, our expert analysis indicates that such cases are uncommon in sub-apps, resulting in few false negatives.

The 2 remaining false negatives in identifying SPO are caused by insufficient contextual information that cannot be directly obtained from subAPI invocations and UI input semantics. For instance, we find that a sub-app collects a file from users by calling wx.uploadFile() and then uploads it to its server, which SPOChecker identifies as a file collection behavior. However, in reality, the sub-app collects a screenshot image (which is also a file) by dynamically specifying the file path, i.e., a file under the screenshot directory. 
In this case, SPOChecker fails to identify the over-collection of the screenshot.

The above analysis indicates that SPOChecker can accurately and comprehensively identify SPO behaviors in sub-apps, and is qualified to perform subsequent measurement analyses.

\section{SPO Measurement}~\label{section:measurement}

In this section, we build a large-scale dataset and conduct a measurement of SPO by answering the following research questions:
\begin{itemize}
    \item \textbf{RQ1. SPO Landscape}. What are the prevalence and characteristics of SPO in the wild?
    \item \textbf{RQ2. SPO Accountability}. Who are the stakeholders in the sub-app ecosystem and what are their responsibilities in the SPO problem? 
    \item \textbf{RQ3. SPO Defense}. What are current privacy protection methods and why they are not enough?
\end{itemize}

By answering these research questions, we make clear the SPO problem in the sub-app ecosystem and get interesting findings and inspiring lessons for further improvement.

\subsection{Dataset}~\label{subsection: dataset}

We use SPOChecker to crawl WeChat sub-apps during Jun 2022.
To make the dataset representative, we remove zombie sub-apps and only keep those used by more than 1,000 different users, indicated by the ``recently used'' field in their metadata.
Eventually, SPOChecker downloads 5,521 sub-apps, covering all common sub-app categories in an official list of mobile app types~\cite{caccat}.

\begin{table}[!h]
\centering
\caption{Our Dataset. }
\label{table:dataset}
\scalebox{1}{
    \begin{tabular}{lll}
    \toprule
    \textbf{\# Sub-apps} & \textbf{\# Valid Privacy Policy }  & \textbf{\# Code Packages} \\
    \midrule
    5,521  & 2,511 & 66,975 \\
    \bottomrule
    \end{tabular}
}
\end{table}

\textbf{Finding 1. Less than half of sub-apps provide valid privacy policies}.
As shown in Table~\ref{table:dataset}, only 2,511 (45.48\%) sub-apps provide valid privacy policies, which is surprisingly low.
For all those 3,010 sub-apps that do not provide valid privacy policies, 749 (24.88\%) have privacy policy texts and links, but these links are either unavailable or linked to blank pages.
The left 2,261 (75.12\%) sub-apps do not provide any privacy policy indicators at all.

\begin{figure}[!h]
    \centering
    \scalebox{0.9}{
        \includegraphics[width=0.45\textwidth]{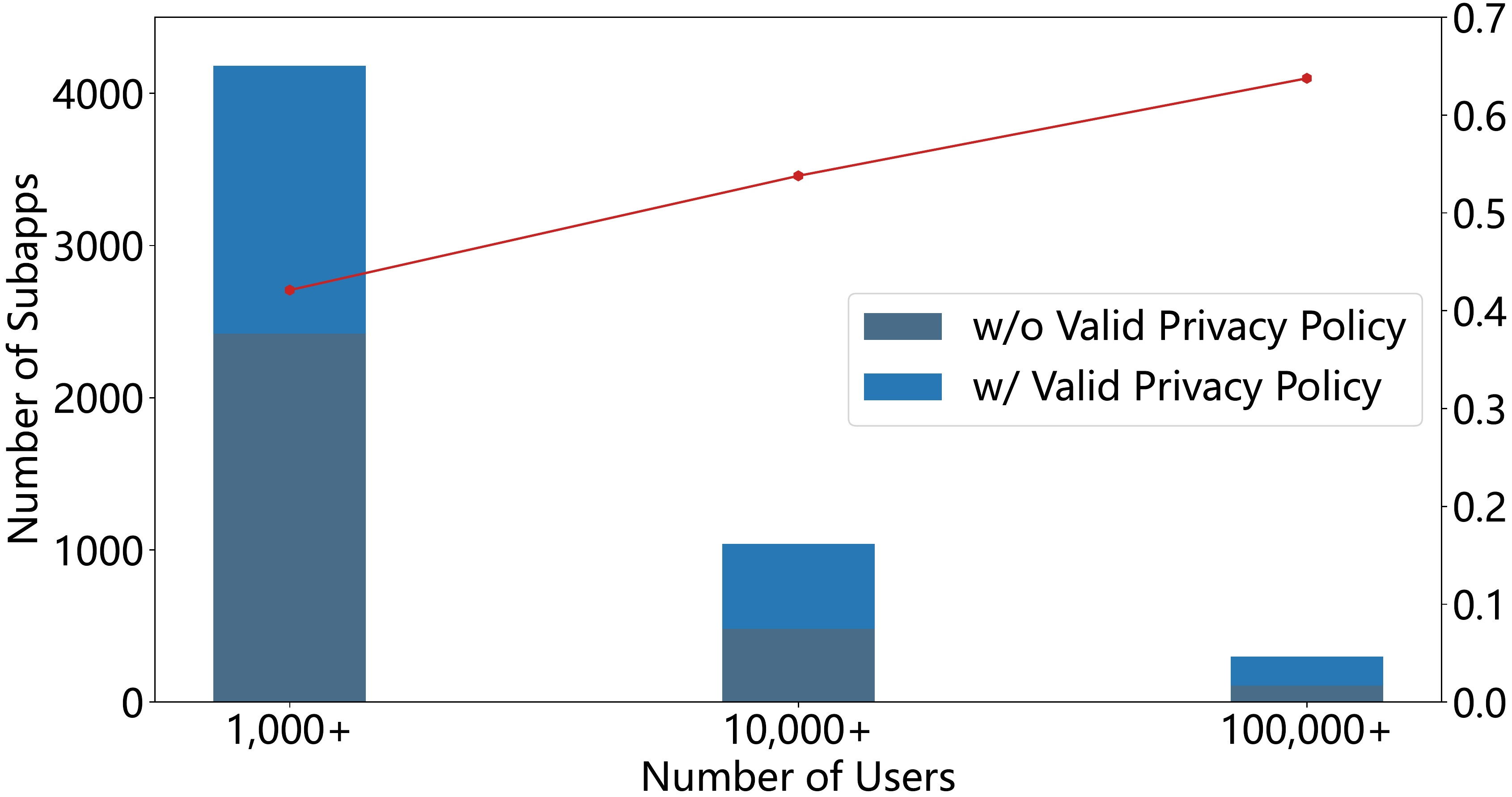}
    }
    \caption{Privacy Policy Providing Rate According to Sub-app Popularity.}
    \label{fig:pp_rate}
\end{figure}

\textbf{Finding 2. Popular sub-apps may have a higher privacy policy providing rate}.
We then study the difference between sub-apps with different popularity, as shown in Figure~\ref{fig:pp_rate}.
We classify sub-apps into three categories according to their ``recently used'' field in the metadata. 
We find that sub-apps with larger ``recently used'' tend to have a higher rate of providing valid privacy policies.
This trend shows that popular sub-apps care more about user privacy.
However, it may also reveal a sad truth that the privacy policy providing rate in the whole ecosystem may be even lower, as there are large amounts of unpopular sub-apps that have less than 1,000 ``recently used''.

\textbf{Finding 3. SubPkgs play an important role in sub-apps}.
Table~\ref{table:dataset} shows that 5,521 sub-apps all together have 66,975 code packages, thus on average each sub-app has 12.13 packages, including 1 mainPkg and 11.13 subPkgs.
We also find that the average size of a subPkg is 5.192MB, while the average size of a mainPkg is 5.371MB, showing that subPkgs provide as much code as mainPkgs do.
Furthermore, we find that there are 35.64\% privacy policies located in subPkgs.
The above results indicate that subPkgs play an important role in sub-apps and considering subPkgs is necessary when studying sub-app privacy.

\begin{tcolorbox} 
\textbf{Lesson Learned:}
The privacy policy providing rate is relatively low in the WeChat sub-app ecosystem.
\end{tcolorbox}

\subsection{SPO Landscape (RQ1)}~\label{subsection: rq1 landscape}

This subsection tries to answer RQ1 by demystifying the prevalence and characteristics of SPO in real-world WeChat sub-apps. 

\subsubsection{SPO Prevalence}~\label{subsubsec: spo prevalence}
We use SPOChecker to detect all SPO behaviors in 2,511 sub-apps.
Note that if a sub-app collects the same privacy item more than once, it is only counted once in our analysis.

\begin{figure}[!h]
  \centering
  \includegraphics[width=0.7\linewidth]{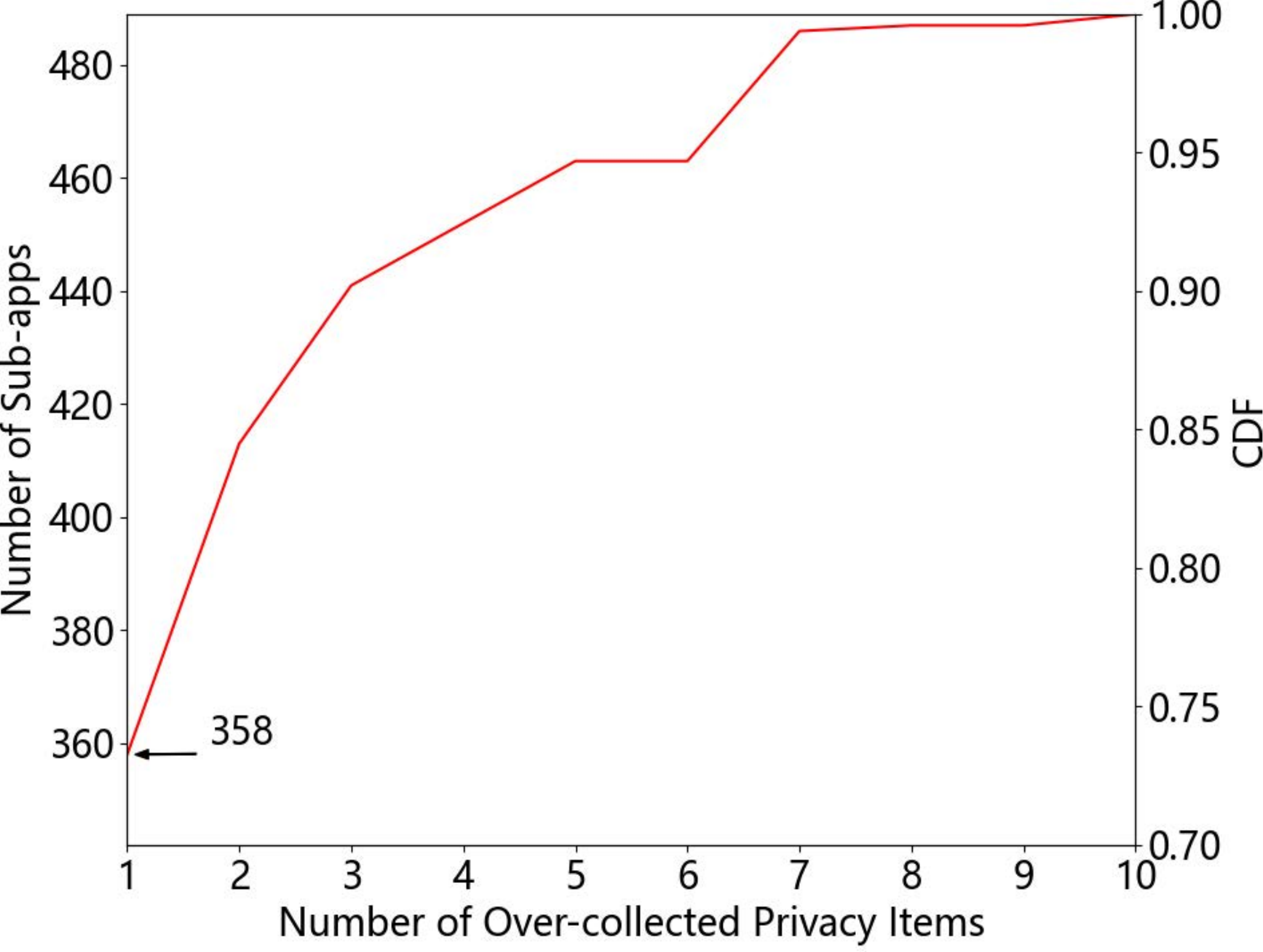}
  \caption{Cumulative Distribution of SPO in 489 Sub-apps. }
  \label{fig:spo}
\end{figure}

\textbf{Finding 4. SPO is very prevalent in the real world.}
In all 2,511 sub-apps, 19.47\% of them (489) contain SPO.
452 (18.00\%) sub-apps have at least 1 but no more than 4 SPO, while 37 (1.47\%) sub-apps have at least 5 SPO.
The most over-collecting sub-apps even over-collect 10 different privacy items.
Note that the result is conservative as we only consider sub-apps with valid privacy policies.
As a comparison, we also list the result of all 5,521 sub-apps in Table~\ref{table:spo_5521}.
For the sub-apps in the dataset that do not provide privacy policies, we consider their privacy collection behaviors all as SPO. 
Once the SPO issues of these sub-apps are taken into account, the overall rate of SPO in the dataset will dramatically increase from 15.65\% to 54.03\%.

\begin{table}[htbp]
\centering
\caption{SPO Result of 5,521 Sub-apps.}~\label{table:spo_5521}
\scalebox{1}{
    \begin{tabular}{ll  ll  ll}
    \toprule
    \textbf{Dataset} & \textbf{\# Collected} & \textbf{\# SPO} & \textbf{SPO\_rate}\\
    \midrule
     2,511 & 4,989 & 781 & 15.65\%\\
     5,521 & 9,153 & 4945 & 54.03\% \\
    \bottomrule
    \end{tabular}
}
\end{table}

\subsubsection{SPO Characteristics}~\label{subsubsec: spo characteristics}
We study SPO in each privacy item and by different categories, and possible objective reasons that lead to SPO.

\begin{figure}[htbp]
  \centering
  \includegraphics[width=0.85\linewidth]{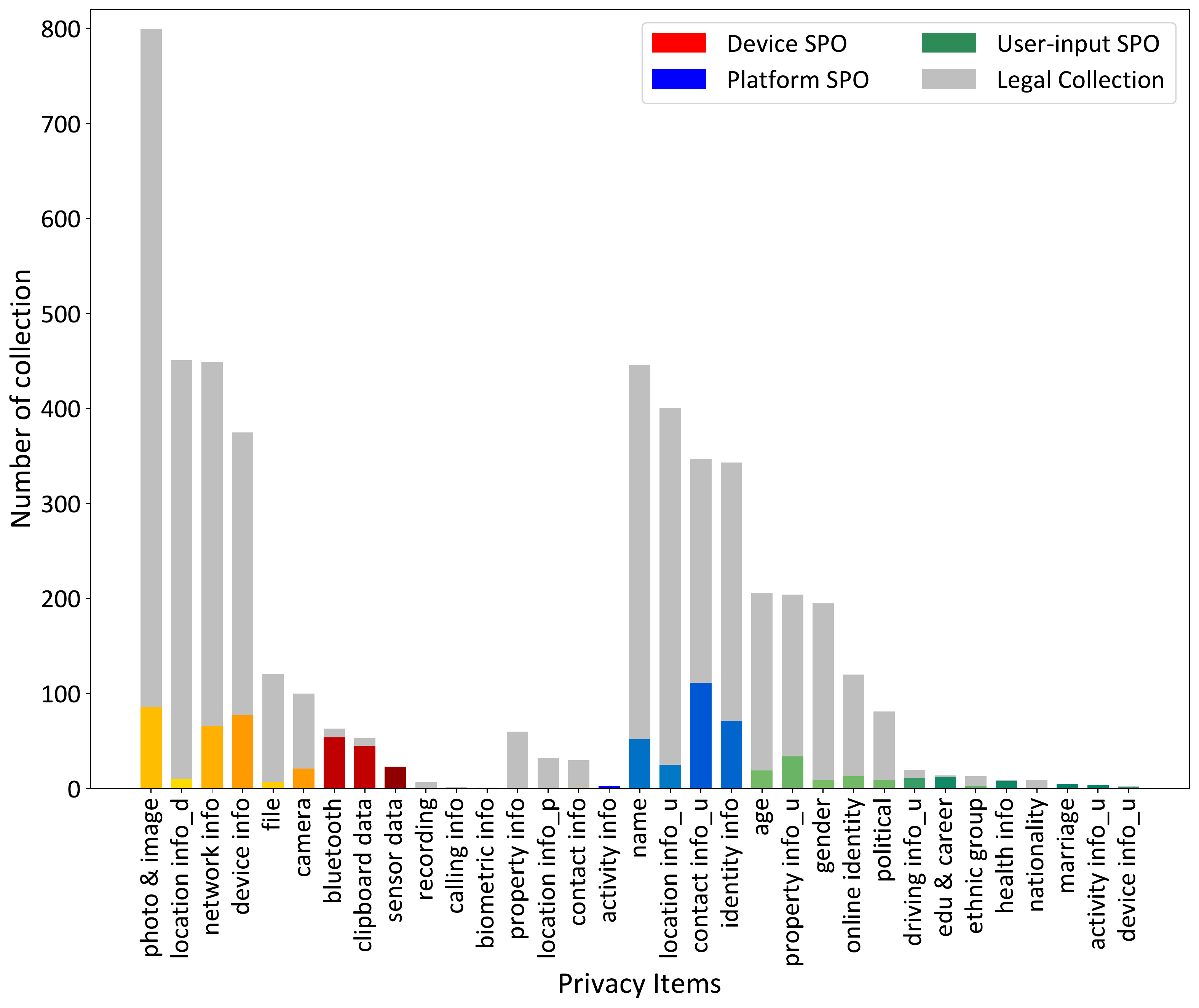}
  \caption{SPO of Different Privacy Items.}
  \label{fig:spo privacy types}
\end{figure}

\textbf{Finding 5: SPO varies in different privacy items, and some privacy items are more over-collected.}
We count the SPO of each privacy item and the results are shown in Figure~\ref{fig:spo privacy types}.
For each privacy item,  the \emph{SPO\_rate} is also calculated, which is the over-collected amount divided by the collected amount and presented in the form of a heatmap in the figure.
As it shows, some privacy items are heavily over-collected. 
The top 10 most over-collected privacy items are shown in Table~\ref{table:top 10 rate}.

\begin{table}[htbp]
\centering
  \caption{TOP 10 SPO Privacy Items and Possible Reasons (PR). }
  \label{table:top 10 rate}
  \scalebox{0.9}{
      \begin{tabular}{llll}
        \toprule
         \textbf{Privacy} & \textbf{Category} & \textbf{SPO\_rate} & \textbf{PR}  \\
        \midrule
        Bluetooth & Device & 85.71\% (54/63) & \ding{172}\ding{174} \\ 
        Clipboard Data & Device & 84.91\% (45/53) & \ding{172}\ding{173}\ding{175} \\
        Contact Info\_u & User-input & 31.99\% (111/347) & \ding{172}\ding{173}\ding{174} \\
        Camera & Device & 21.00\% (21/100) & \ding{172}\ding{173}\\
        Identity Info & User-input & 20.70\% (71/343) & \ding{172}\ding{173}\ding{174}\\
        Device Info & Device & 20.53\% (77/375) & \ding{172}\ding{173}\\
        Property Info\_u & User-input & 16.67\% (34/204) & \ding{172}\ding{173}\ding{174} \\
        Network Info & Device & 14.70\% (66/449) & \ding{172}\ding{173}\\
        Name & User-input & 11.66\% (52/446) & \ding{172}\ding{173}\\
        Political & User-input & 11.11\% (9/81) & \ding{172}\ding{173}\ding{174} \\
        \bottomrule
        \end{tabular}
    }
\end{table}

We then manually look into these top 10 over-collected privacy items and summarize possible reasons. 
Initially, all SPO behaviors can be subjectively used by sub-apps for \textit{Covert Data Harvesting} (\ding{172}). 
Nevertheless, we try to infer the objective reasons leading to SPO by studying the specific code usage and privacy policy statements.
As a result, we summarize the following possible objective reasons:

\textit{Vague Statement} (\ding{173}). 
The primary reason for some privacy items to be over-collected is an imprecise and coarse-grained statement in sub-apps privacy policies.
For example, developers only claim to collect ``personal information such as name, ID, etc'' without explicitly specifying the marriage and political information they collected.
For another example, sub-apps may use \emph{wx.chooseInvoice()} to get users' invoices which have their organization and consumption information, but only state collecting ``financial information'' instead of ``invoices''.
More examples are shown in Table~\ref{table:vague statement} in Appendix.

\textit{Privacy Unawareness} (\ding{174}). 

This may be because developers only focus on the functionality of subAPIs but are not aware of the privacy risks behind them. 
For example, the usage of Bluetooth and screenshots may leak more user privacy than they expected, e.g. the \emph{UUID} field of Bluetooth may be used to get the user's location under some circumstances~\cite{bluetoothUUID}, and screenshots may leak time and system information. 
Developers unaware of such risks will not state these risks in their privacy policies.

\textit{Perception Evolution} (\ding{175}). 
The concept of privacy on mobile platforms keeps evolving, where some items not considered privacy data before may be seemed as privacy data as people's perceptions change. 
For example, Google does not restrict the access to user's clipboard data until Android 10~\cite{android10}. 
The restrictions on visiting the clipboard thus lagged on super-apps, and sub-apps also do not make clear of this in their privacy policies.

\textbf{Finding 6: Vague statement, privacy unawareness, and perception evolution are objective reasons leading to SPO}.
Therefore, we recommend that sub-app developers focus on these objective issues to provide more accurate descriptions of privacy policies, ensuring that users have a comprehensive understanding of all privacy collection behaviors. 
In addition, for items that may contain hidden or indirect privacy information, such as Bluetooth UUID, developers should take measures to prevent the collection of such hidden information (if hidden information is not used), or clearly state the purpose of this privacy information in the privacy policies.


\begin{table}[htbp]
\centering
\caption{SPO by Category.}
\label{table:spo types}
\scalebox{1}{
    \begin{tabular}{lll}
    \toprule
    \textbf{Category} & \textbf{\# Collected }  & \textbf{\# SPO} \\
    \midrule
    Device Privacy & 2,444 & 389 (15.92\%) \\
    Platform Privacy & 125 & 4 (3.2\%)  \\
    User-input Privacy & 2,420 & 388 (16.03\%)  \\
    Sum & 4,989 & 781 (15.65\%)  \\
    \bottomrule
    \end{tabular}
}
\end{table}

\textbf{Finding 7: Platform privacy is the least over-collected category. }
We also study SPO of different privacy categories, as shown in Table~\ref{table:spo types}.
These 2,511 sub-apps collect a total of 4,989 privacy items, while 781 (15.65\%) of them are over-collected. 
The over-collection rates for device and user-input privacy are roughly the same. 
Platform privacy is the least over-collected, which can be more easily monitored and constrained by the WeChat platform.

In contrast, user-input privacy is more diverse and hard to be constrained. 
For example, the SPO\_rates for \emph{contact\_p}, \emph{contact\_u} are 3.33\% and 31.99\% respectively, where contact information is significantly more over-collected through the way of user-input than calling platform subAPIs.


\begin{tcolorbox} 
\textbf{Lesson Learned:}
SPO is prevalent in the real world, and privacy items that are more diverse and hard to regulate tend to be more over-collected. 
Sub-app developers should provide more precise and comprehensive descriptions to prevent unintentional SPO.

\end{tcolorbox}

\subsection{SPO Accountability (RQ2)}~\label{subsection:rq2 accountability}

\begin{figure}[htbp]
    \centering
    \includegraphics[width=0.45\textwidth]{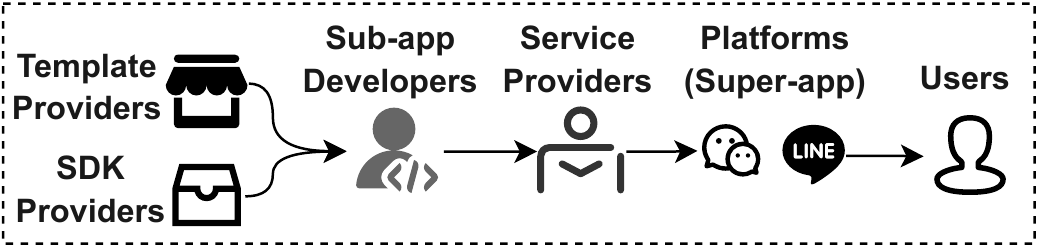}
    \caption{Stakeholders in Sub-app Ecosystem.}
    \label{fig:stakeholders}
\end{figure}

This subsection tries to answer RQ2 by identifying stakeholders involved in this ecosystem and their respective responsibilities regarding user privacy protection.
First, through an in-depth study on real-world sub-apps, we summarize important stakeholders in developing and operating sub-apps, as shown in Figure~\ref{fig:stakeholders}.

In this ecosystem, the \emph{Super-apps} set up the platform and build a bridge between sub-apps and users.
The \emph{Service Providers} operate the sub-apps, claim what privacy they want in the privacy policies and collect them through code and UI interaction, to provide specific services.
The development of sub-apps may be done by service providers themselves, or outsourced to specialized \emph{Sub-app Developers}.
As such, service providers are directly responsible for protecting user privacy, while super-apps should regulate sub-apps and provide support to service providers, developers, and users in privacy protection.

Furthermore, we find templates and SDKs are widely used in developing sub-apps. 
\emph{Templates}, or sub-app generators,  are those general sub-app frameworks that can generate different sub-apps with only a few changes and configurations. 
Consequently, if privacy is not well-protected in one template, it may affect a large number of downstream sub-apps. 
The same problem also applies to \emph{SDKs}, which make parts of a sub-app. 

In this subsection, we study privacy collection and over-collection in sub-app templates and SDKs.

\subsubsection{Detecting Templates and SDKs in Sub-apps}~\label{subsubsec: detecting templates and SDKs}
According to our observation, sub-apps generated by the same template have almost identical layouts and code, with only slight differences in configuration and resource files, such as the service name, product description, and images. 
Therefore, we propose a two-level clustering algorithm (Algorithm~\ref{alg:detecting_sub-app_templates}) that groups similar sub-apps into the same cluster, by first comparing page structure and then page content. 
A similar idea is also applied to clustering sub-app SDKs.
We discuss the details below. 

\begin{algorithm}[htbp]
    \caption{Detecting Sub-app Templates} \label{alg:detecting_sub-app_templates}
    \begin{algorithmic}[1] 
        \Require Sub-app set $S$, where $s.rt$, $s.ctn$, $s.dev$ denote page route, file content and developer of sub-app $s$; threshold $\theta_1,\theta_2$. 
        \State Template set $T = \emptyset$  
        \For{each $s$ in $S$}
            \For{each $t$ in $T$}
                \If{$Sim(s.rt, t.rt) \geq \theta_1 $ \&  $Sim(s.ctn, t.ctn) \geq \theta_2$}
                    \State Add $s$ to $t$
                \EndIf
            \EndFor
            \If{$s$ not added to any $t$}
                \State Add New $t$($s$) to $T$
            \EndIf
        \EndFor
        \For{$t$ in $T$ \textbf{where} ( $\|t\| < 2$ ) or ( $\|t.dev\| < 2$)}
            \State Remove $t$ from $T$
        \EndFor
        \State Return $T$
    \end{algorithmic}
\end{algorithm}

As shown in Algorithm 1, to detect sub-app templates, we first cluster sub-apps with the same page routes into different groups. 
Files of each sub-app from the same group are compared one by one, and their similarity is calculated. 
The similarity results are either very close to 1 or obviously below 0.9. 
Therefore, the similarity thresholds are set to 0.9, and sub-apps with similarity above the threshold are classified into one template. 
We then remove templates with only one sub-app or one developer.

The SDK detection algorithm is similar. 
First, all JavaScript files in sub-app packages with the same names are clustered into groups, and files in the same group are compared for similarity. 
Statistical analysis shows that SDK file similarity tends to equal 1 or converge to 1, so a relatively high threshold of 0.95 is set. 
Subsequently, we consider that SDK files tend to be used by various sub-apps, so we also count the number of file occurrences among different sub-apps, and only keep files used by more than 100 sub-apps as the candidate SDK files.

We then merge these files into SDKs based on their path and obtain the final SDK list. Finally, we manually confirmed the accuracy of the template and SDK identification.

\begin{table}[htbp]
\centering
\caption{Detected Sub-app Templates and SDKs in 5,521 Sub-apps.}~\label{table:templates and sdks}
\scalebox{1}{
    \begin{tabular}{l l l}
    \toprule
    \textbf{Detected} & \textbf{Count } & \textbf{\# Sub-apps } \\
    \midrule
    Templates & 272 & 1,307 (23.7\%)  \\
    SDKs & 307 & 3,430 (62.1\%)  \\
    \bottomrule
    \end{tabular}
}
\end{table}

\textbf{Finding 8: Templates and SDKs are heavily used in sub-apps. }
As shown in Table~\ref{table:templates and sdks}, for all 5,521 sub-apps in our dataset, 1,307 (23.7\%) sub-apps are generated by 272 distinct templates, and 3,430 (62.1\%) sub-apps use 307 distinct SDKs. 
It should be noted that the rates presented here are conservative, as our detection method ignores less popular templates and SDKs, such as templates only used by one sub-app. 
Table~\ref{table:top10temp} and Table~\ref{table:top10sdk} list the top 10 templates and SDKs, respectively. 
We observed that popular templates include those related to online shopping, online ordering, and recommendation, while the most commonly used SDKs are those that provide UI components, parsing, and network communication functions.
These business models and functions are crucial to sub-apps, often involving their core services. 
Therefore, the study of SPO in templates and SDKs is worth exploring.

\begin{table}[htbp]
\centering
    \caption{Top 10 Sub-app Templates. }
    \label{table:top10temp}
    \scalebox{0.9}{
        \begin{tabular}{llcl} 
        \toprule
        \textbf{Rank} & \textbf{Template} & \textbf{\# Sub-apps} & \textbf{Function}\\ \midrule 
        1 & \textbf{XX Travelling} & 69 & Bicycle-sharing\\
        2 & \textbf{Hualala} & 56 & Catering \& Takeaway \\
        3 & \textbf{Wuuxiang\_v1} & 39 & Catering \& Takeaway \\
        4 & \textbf{Youzan\_v1} & 37 & Online Shopping  \\
        5 & \textbf{Biaodian\_v1} & 33 & Online Shopping\\
        6 & \textbf{mini-Video} & 32 & Video Recommend \\
        7 & \textbf{Youzan\_v2} & 31 & Online Shopping \\
        8 & \textbf{Court} & 26 & Online Court  \\
        9 & \textbf{Wuuxiang\_v2} & 26 & Catering \& Takeaway \\
        10 & \textbf{Biaodian\_v2} & 22 & Online Shopping \\
        \bottomrule
        \end{tabular}
    }
\end{table}

\begin{table}[htbp]
\centering
    \caption{Top 10 Sub-app SDKs. Note that One Sub-app May Use Multiple SDKs. }
    \scalebox{0.9}{
        \begin{tabular}{llcl} 
        \toprule
        \textbf{Rank} &\textbf{SDK} & \textbf{\# Sub-apps} & \textbf{Function} \\ \midrule 
        1 & \textbf{Vant-Weapp} & 1,252 & UI comp. \\
        2 & \textbf{wxParse} & 815 & parsing \\
        3 & \textbf{uniapp} & 626 & front-end   \\
        4 & \textbf{uview-ui} & 583 & WebUI comp.  \\
        5 & \textbf{mp-html} & 462 & rich text   \\
        6 & \textbf{trtc-room} & 239 &  RTC SDK   \\
        7 & \textbf{ext-player} & 230 & RTC SDK  \\
        8 & \textbf{WeUI} & 201 & UI framework   \\
        9 &  \textbf{ThorUI} & 199 & UI comp. \\
        10 & \textbf{Youzan-SDK} & 201 & framework \\
        \bottomrule
        \end{tabular}
    }
    \label{table:top10sdk}
\end{table}

\subsubsection{SPO in Templates and SDKs}~\label{subsubsec: SPO in templates and SDKs}
We study the difference between sub-apps generated by templates and those not by templates (\emph{non-templates} for short), as shown in Table~\ref{Table:templates VS non-templates}.

\begin{table}[htbp]
\centering
\caption{SPO of Templates and Non-templates in 2,511 Sub-apps With Valid Privacy Policies.}
\label{Table:templates VS non-templates}
\scalebox{1}{
    \begin{tabular}{l l l l}
    \toprule
    & \textbf{\# Sub-apps } & \textbf{\# Collected } & \textbf{\# SPO} \\
    \midrule
    Templates     &   564 & 1,693 & 277 \\
    Non-Templates & 1,947 & 3,296 & 504 \\
    \bottomrule
    \end{tabular}
}
\end{table}

\textbf{Finding 9: SPO in templates is more severe than in non-templates. }
In all 2,511 sub-apps with valid privacy, 564 sub-apps are generated by templates, while 1,947 are non-templates. 
Overall, sub-apps generated from templates collect and over-collect significantly more user privacy than non-template sub-apps.
As shown in Table~\ref{Table:templates VS non-templates}, non-template sub-apps collect an average of 1.69 privacy items, whereas template sub-apps collect 3.00. 
In terms of SPO, non-template sub-apps have 0.26 SPO per sub-app, while template ones 0.49. 
Furthermore, out of 21 privacy items collected more than 10 times, 14 (66.67\%) of them have a higher SPO\_rate in template sub-apps than in non-template sub-apps.

\begin{table}[htbp]
\centering
\caption{Collection and SPO Distribution in 1,593 Sub-apps With Valid Privacy Policies and SDKs.}
\label{Table:sdk-nonsdk}
\scalebox{1}{
    \begin{tabular}{l l l l}
    \toprule
    & \textbf{\# Collection } & \textbf{\# SPO } & \textbf{Size (MB)} \\
    \midrule
    Sub-apps & 1,443 & 176 & 10.563 \\
    SDK code & 216 & 14 & 0.31 \\
    Ratio & 14.97\% & 7.95\% & 2.93\% \\
    \bottomrule
    \end{tabular}
}
\end{table}

\textbf{Finding 10: SDKs play an important role in the privacy collection of sub-apps.}
In all 2,511 sub-apps with valid privacy policies, 1,593 sub-apps use at least one SDK, and we study the location of SPO in these sub-apps.
Note that as SDKs usually do not have specific Web pages, we do not count user-input privacy here.
As shown in Table~\ref{Table:sdk-nonsdk}, out of the 1,593 sub-apps, a total of 1,443 privacy items are collected, with 216 (14.97\%) of them located in SDK files. 
Furthermore, these sub-apps over-collected 176 privacy items, with 14 (7.95\%) of them being over-collected by code belonging to SDKs. 
Additionally, it should be noted that the code size of SDK files is small, accounting for only 2.93\% (0.31MB out of 10.563MB - see \S~\ref{subsection: dataset}) of all code.

Therefore, we can conclude that the importance of sub-app SDKs in protecting user privacy should not be ignored.

\begin{tcolorbox} 
\textbf{Lesson Learned: }
Sub-app service providers should be responsible for user privacy. 
Also, template and SDK developers, as well as the regulation of templates and SDKs, play an important role in sub-app privacy protection.
\end{tcolorbox}

\subsection{SPO Defense (RQ3)}~\label{subsection: rq3 defense}

In this research question, we study the current privacy-related defense methods in the WeChat platform and discuss why they are not enough.

\subsubsection{Current Defense Methods}~\label{subsubsec: current defense methods}
Unlike Android and iOS mobile platforms, the WeChat platform does not mandate sub-app developers to provide privacy policies, nor does it check the privacy policies in sub-apps. 
As a result, the privacy policy providing rate is relatively low in WeChat sub-apps.
Table~\ref{table:spo_5521} proves sub-apps without valid privacy policies tend to collect and over-collect more privacy items.
Nevertheless, in this subsection, we discuss the existing protection methods adopted by WeChat below.

\textbf{Finding 11: Existing protection methods are unable to force sub-apps to protect users' privacy}.
Currently, WeChat provides two methods to protect user privacy: 1) a permission mechanism, and 2) privacy protection guidelines. 
However, neither of these two methods is mandatory, and sub-apps may bypass them.

\textit{Permission Mechanism}. 
Like in high-version Android, WeChat requires sub-apps to apply for permissions and dynamically prompts a window to inform users when using certain subAPIs, but it only covers a small set of privacy items, as listed in WeChat official documents~\cite{scope}. 
Specifically, for all 125 subAPIs listed in Table~\ref{table:taxonomy}, only 43 of them (34.4\%) need explicit permission grants.
For example, \textit{wx.getConnectedWifi()} can return users' network information including BSSID, but using it does not need any permission in WeChat.
Furthermore, this subAPI-based method cannot be applied to user-input privacy items, which are collected from the UI instead of subAPIs. 
As a result, the permission mechanism is not enough to protect all user privacy.

\begin{figure}[htbp]
    \centering
    \includegraphics[width=0.48\textwidth]{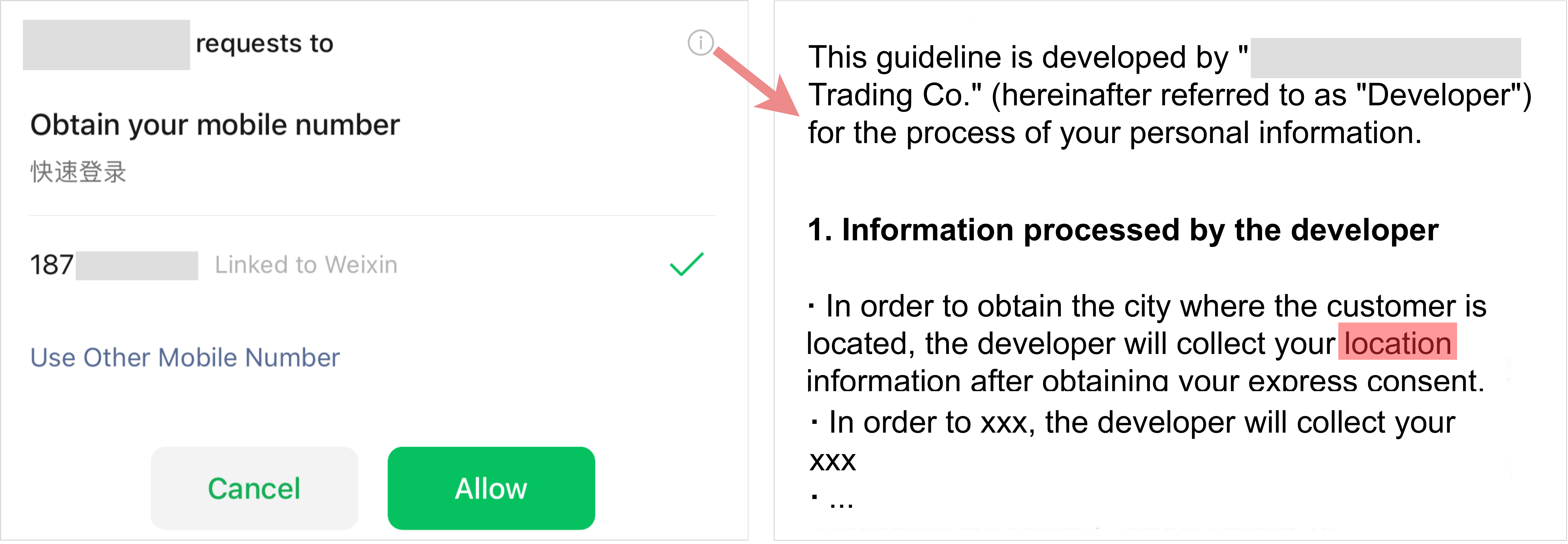}
    \caption{WeChat Permission and Privacy Protection Guidelines.}
    \label{fig:wechatguide}
\end{figure}

\textit{Privacy Protection Guidelines}.
WeChat also requires developers to fill in a form, named privacy protection guidelines~\cite{wechatguide}, to make clear what privacy is collected. 
The basic framework of the form, which specifies the privacy items that need to be clarified, is generated by WeChat through code auditing. Subsequently, sub-app developers are responsible for filling in the form with the appropriate details.
However, our experiments find that this form is not rigorously checked, thus developers may provide incomplete information. 
For example, in our dataset with 5,521 sub-apps, 286 (5.18\%) of them submit a blank guideline with no privacy collection statements.
In addition, the entry of the form, as shown in Figure~\ref{fig:wechatguide}, located in a corner of a pop-up box, is not easy for users to find, thus weakening its effectiveness.

\subsubsection{Evaluating Existing Protection.}~\label{subsubsec: evaluating existing protection}
We also investigate whether existing methods can promote privacy protection.

First, we try to quantify the impact of existing protection on SPO.

We find that both the permission mechanism and the code auditing of privacy protection guidelines fail to cover all collection behaviors of one privacy item. 
For example, WeChat has a permission named \textit{userLocation} for sub-apps to apply for user's location, but only the invocations of 4 subAPIs, i.e. getLocation(), chooseLocation(), startLocationUpdate() and startLocationUpdateBackground(), need to request this permission.
However, there are 10 subAPIs (see Table \ref{table:taxonomy}) that can be used by sub-apps to acquire user location.
As a result, location privacy is only partially protected by existing protection methods.

\begin{table}[htbp]
\centering
    \caption{The SPO\_rate of Privacy Items under Different Levels of Protection.}
    \label{table:guideitem}
        \begin{tabular}{llll} 
        \toprule
        \textbf{Protection Level} &\textbf{\# Collect} & \textbf{\# SPO} & \textbf{SPO\_rate} \\ \midrule 
        Not Protected & 3,472 & 607 & 17.48\% \\
        Partial Protected & 1,420 & 171 & 12.04\% \\
        Fully Protected & 97 & 3 & 3.09\% \\
        \bottomrule
        \end{tabular}
\end{table}

Therefore, we classify privacy items into three protection levels based on the extent to which they are protected by one of the existing two protection methods: \textit{Not Protected}, \textit{Partially Protected}, and \textit{Fully Protected}.
After that, we calculate the SPO\_rate of privacy items under different levels of protection, as shown in Table~\ref{table:guideitem}.

\textbf{Finding 12. Existing protection methods can significantly reduce the SPO\_rate, but their main drawback is that the coverage is not complete.}
We can see from the table that the better a privacy item is protected, the lower its SPO\_rate.

Therefore, one way to improve user privacy protection is to increase the coverage rate of existing protection methods.

\begin{tcolorbox} 
\textbf{Lesson Learned: }
Super-apps can adopt more complete and visible protection methods and stricter regulations on developers to better protect user privacy.

\end{tcolorbox}


\section{Limitations \& Discussions}~\label{section:discussion}

\textbf{Detecting Privacy Collection.} 
SPOChecker relies on static analysis on JavaScript to identify SPO and therefore is subject to the limitations of static analysis tools.
However, we have taken the following measures to ensure the reliability of SPOChecker:
First, we make sure the invocation of subAPIs does have data transmission by carefully selecting \textit{Source} subAPIs (Section~\ref{subsection:detecting collection}).
Second, we apply sentimental analysis to privacy policies. 
As a result, if a sub-app only uses privacy items locally without sending them to its server, and it clearly states the local usage in its privacy policy, SPOChecker can exclude such behaviors from SPO (\S~\ref{subsection: identifying SPO}).
Finally, our evaluation with SPOChecker (\S ~\ref{subsec:evaluating}) proves its effectiveness and accuracy. 
Nevertheless, we agree that sophisticated static analysis works~\cite{stein2019static, park2018static, staicu2020extracting} can further improve SPOChecker.

\textbf{Identifying User-input Privacy.} 
We recognize user-input privacy by matching keywords in texts from components such as labels and tips. 
However, only considering texts may be incomplete, as non-text resources may also contain privacy.
For example, images showing males or females with a toggle button can also be a way to collect users' gender information.
We do not find such cases in our dataset, and we plan to look into this problem in our future work.

\textbf{Other Sub-app Ecosystems. } 
In this work, we focus on privacy over-collection in the WeChat sub-app ecosystem, while several other super-apps support the app-in-app paradigm.
We choose WeChat as the study target because it provides the largest and most representative sub-app ecosystem yet~\cite{zhang2022identity, zhang2021measurement}. 
According to our observations, most sub-app ecosystems have similar architecture and technology as WeChat to support Web-based sub-apps, thus methods used in our study can also be applied to other platforms.

\section{Related Work}~\label{section:related work}

\textbf{App-in-app Security}.
The app-in-app paradigm is very popular now and several work~\cite{zhang2018empirical, zhang2019app, liu2020industry, lu2020demystifying, zhang2021measurement, zhang2022identity} have studied its security. 
Most of them~\cite{zhang2018empirical, zhang2019app, lu2020demystifying, zhang2022identity} focus on the security issues such as access control between the native code and Web code. 
Y. Zhang et al~\cite{zhang2021measurement} conducts a preliminary study on code properties using millions of WeChat sub-apps. 
They propose a method to download sub-apps but subPkgs, which make up an important part of the sub-apps shown in our work, are not included.
H. Lu et al~\cite{lu2020demystifying} recognize subAPIs that are not well protected by super-apps and can be used to illegally access system resources. 
We optimize their method with high-quality fuzzing seeds and find more subAPIs that can access user privacy. 
Recently, more and more attention has been paid to the vulnerabilities and bugs present in sub-apps.
For example, Y.Yang et al\cite{crossminiapp} study cross-sub-app request forgery and possible consequences, while 
T.Wang et al\cite{miniappbug} present an empirical study on WeChat sub-app bugs caused by common JavaScript errors such as type errors.
These works show that security and privacy in the sub-app ecosystem are of great importance. 
However, none of the above work focuses on privacy over-collection in sub-apps, and ours is the first of its kind.



\textbf{Web Privacy}.
Privacy on websites is a classic topic of security research. 
A.R.A. Bouguettaya et al~\cite{bouguettaya2003privacy} discuss the definition of online privacy, the possible ways websites collect privacy, and how to preserve user privacy. 
G. Yee et al~\cite{yee2004semi} provide semi-automated approaches to derive privacy policy based on community consensus and existing privacy policies. 
With the help of machine learning, S. Zimmeck et al~\cite{zimmeck2014privee} implement a classifier to recognize the privacy usage of websites, while T. Libert~\cite{libert2018automated} presents a large-scale audit of a million website privacy policies and provides a thorough analysis.
Sub-apps, as a special form of Web apps, share similar privacy problems but also face new challenges. 
This paper fills a gap in Web privacy research by studying a new form of Web apps.

\textbf{Privacy Policy Analysis}.
There are many work analyzing privacy policy~\cite{andow2019policylint, zimmeck2016automated, slavin2016toward, wang2018guileak, yu2018ppchecker, andow2020actions, nan2015uipicker, nan2018finding, zhang2019app}. 
For example, B. Andow et al~\cite{andow2019policylint} find internal semantic contradictions in the privacy policies of mobile apps. 
S. Zimmeck et al~\cite{zimmeck2016automated}, R. Slavin et al~\cite{slavin2016toward}, L. Yu et al~\cite{yu2018ppchecker} reveal the inconsistency between code behavior and privacy policy, and develop tools to help developers and users to detect such inconsistency. 
Some work~\cite{nan2015uipicker, huang2015supor} focus on the problem of user-input privacy over-collection in mobile apps, based on which X. Wang et al~\cite{wang2018guileak} study the inconsistency between user-input privacy and privacy policy in mobile apps. 
Different from the above works, this paper focuses on the sub-app over-collection problem, where a sub-app collects more than it claims in its privacy policy.

\section{Conclusion}~\label{section:conclusion}

In this paper, we perform the first systematic study of privacy over-collection in WeChat sub-apps. 
We propose \emph{SPOChecker}, a framework that can automatically collect sub-apps and detect SPO in the wild. 
Using SPOChecker, we collect 5,521 popular WeChat sub-apps and study the problem from three aspects, including the landscape, accountability, and defense methods of SPO.
We reveal the seriousness of privacy over-collection in the current WeChat sub-app ecosystem, uncover the possible reasons and responsible parties, and discuss the deficiencies of existing defense methods and the directions for improvement. 
We hope our work can help the community to better understand SPO and protect privacy in sub-apps. 

\bibliographystyle{ACM-Reference-Format}
\bibliography{main}

\appendix

\section{Vague Statements in Privacy Policies}

We list some common examples of vague statements in Table~\ref{table:vague statement}.

\begin{table}[!h]
\centering
  \caption{Examples of Vague Statement.}
  \label{table:vague statement}
  \scalebox{1}{
      \begin{tabular}{lp{3.3cm}p{3.3cm}} 
        \toprule
        \textbf{No.} & \textbf{Vague Statement} & \textbf{Collected Privacy Items} \\
        \midrule
        1& ``We will collect your personal health and physical information for...'' & WeChat run data, disease history, blood type, height, weight\\
        2& ``We will collect information about your transactions/order information, including but not limited to the amount of the transaction, the recipient name of the goods, etc.'' & property information, tax number, tax organization name, recipient address, recipient phone number \\
        3& ``For real name verification, we will obtain camera access and collect relative information about you'' & camera, biometric information (face), name, ID, other personal documents \\
        \bottomrule
        \end{tabular}
    }
\end{table}

\end{document}